\documentclass[acmsmall,screen,nonacm]{acmart}
\AtBeginDocument{%
  }

\setcopyright{acmlicensed}
\copyrightyear{2018}
\acmYear{2018}
\acmDOI{XXXXXXX.XXXXXXX}
\acmConference[Conference acronym 'XX]{Make sure to enter the correct
  conference title from your rights confirmation email}{June 03--05,
  2018}{Woodstock, NY}
\acmISBN{978-1-4503-XXXX-X/2018/06}

\usepackage{tabularx}
\usepackage{multirow}
\usepackage{array}
\usepackage{bm}
\usepackage{cleveref}
\usepackage{caption}
\usepackage[linesnumbered,ruled,vlined]{algorithm2e}
\SetAlFnt{\small}
\SetAlCapFnt{\small}

\begin{document}

\title{Blockchain Application in Metaverse: A Review}

\author{Bingquan Jin}
\authornote{Both authors contributed equally to this research.}
\author{Hailu Kuang}
\authornotemark[1]
\email{hailukuang@hainanu.edu.cn}
\affiliation{%
  \institution{Hainan University}
  \city{Haikou}
  \country{China}}

\author{Xiaoqi Li}
\affiliation{%
  \institution{Hainan University}
  \city{Haikou}
  \country{China}}
\email{csxqli@ieee.org}

\renewcommand{\shortauthors}{Jin et al.}

\begin{abstract}
  In recent years, the term Metaverse emerged as one of the most compelling concepts, captivating the interest of international companies such as Tencent, ByteDance, Microsoft, and Facebook. These company recognized the Metaverse as a pivotal element for future success and have since made significant investments in this area. The Metaverse is still in its developmental stages, requiring the integration and advancement of various technologies to bring its vision to life. One of the key technologies associated with the Metaverse is blockchain, known for its decentralization, security, trustworthiness, and ability to manage time-series data. These characteristics align perfectly with the ecosystem of the Metaverse, making blockchain foundational for its security and infrastructure. This paper introduces both blockchain and the Metaverse ecosystem while exploring the application of the blockchain within the Metaverse, including decentralization, consensus mechanisms, hash algorithms, timestamping, smart contracts, distributed storage, distributed ledgers, and non-fungible tokens (NFTs) to provide insights for researchers investigating these topics.
\end{abstract}





\keywords{Blockchain, Metaverse, Decentralization}


\maketitle

\section{INTRODUNCTION}

Since the initial proposal of the Metaverse, it has been envisioned as a desirable realm. However, the year 2021 marked a pivotal moment as numerous enterprises began to engage with technologies such as virtual reality (VR), extended reality (XR), 5G, artificial intelligence (AI), blockchain, and cloud computing. This surge in interest has triggered a wave of enthusiasm surrounding the Metaverse, establishing 2021 as a landmark year for its development. Numerous Internet companies have invested heavily in the Metaverse, demonstrating a strong commitment to establishing a robust Metaverse ecosystem and showcasing its immense potential. However, a major hurdle remains: the foundational technologies necessary for the Metaverse are still underdeveloped. Despite rapid advancements in science and technology, current capabilities are insufficient to support a fully functional Metaverse, necessitating continued progress and breakthroughs.

Globally, various industries are engaging in discussions about the Metaverse. They have identified six supporting technologies essential to its development: Interactivity, Games, Artificial Intelligence (AI), Blockchain, Networking and Computing, and the Internet of Things (IoT). 
Among these, blockchain stands out as a fundamental technology for the Metaverse ecosystem, often described as the bridge linking the real and virtual worlds. It provides the necessary framework for equivalence mapping and conversion between these realms.

Since the inception of Bitcoin in 2008, blockchain has evolved significantly, incorporating various advancements and leading to practical applications across multiple sectors. Ongoing research delves deeper and expands into broader areas, keeping blockchain at the forefront of technological innovation.
Blockchain possesses characteristics such as decentralization, openness, independence, and security, making it well-suited to meet the needs of various stakeholders within the Metaverse ecosystem. Its applications span multiple aspects of this ecosystem. Key elements of blockchain—including decentralization, consensus mechanisms, hash algorithms, timestamps, distributed storage, distributed ledgers, and non-fungible tokens—serve as a robust infrastructure for the Metaverse. Furthermore, these components represent the fundamental technologies necessary for the effective functioning of the Metaverse ecosystem and play a significant role in contemporary research on the subject.

Among the various technologies being researched, the application of blockchain in the Metaverse is advancing rapidly. Current research focuses primarily on decentralization, consensus mechanisms, hash algorithms, timestamp, smart contracts, distributed storage, distributed ledgers, and non-fungible tokens (NFTs). Companies such as Facebook, while primarily investigating VR equipment and social platforms, also recognize the importance of blockchain applications. Similarly, Robiox conducts research within blockchain alongside its work in game engines, while organizations like IBM and Ava Labs are exploring blockchain applications within the Metaverse ecosystem. These enterprises are engaged in foundational research about the decentralization layer of the Metaverse and are establishing the ecological infrastructure necessary for its development.

In summary, the contributions of this paper are as follows:
\begin{itemize}
  \item We provide a comprehensive overview of blockchain applications in the Metaverse by examining foundational components, including decentralization, consensus, hash, timestamps, smart contracts, distributed storage, distributed ledgers, and non-fungible tokens (NFTs).
  \item We analyze each technology individually, assess its role and implementation within the Metaverse ecosystem, and evaluate the current state of these applications.
  \item  We further explore how blockchain technologies are specifically aligned with the needs of the Metaverse, investigate their interconnections, and outline future research directions.
\end{itemize}

\begin{figure}[htbp]
  \centering
  \includegraphics[width=0.8\textwidth, trim=10 1 10 10, clip]{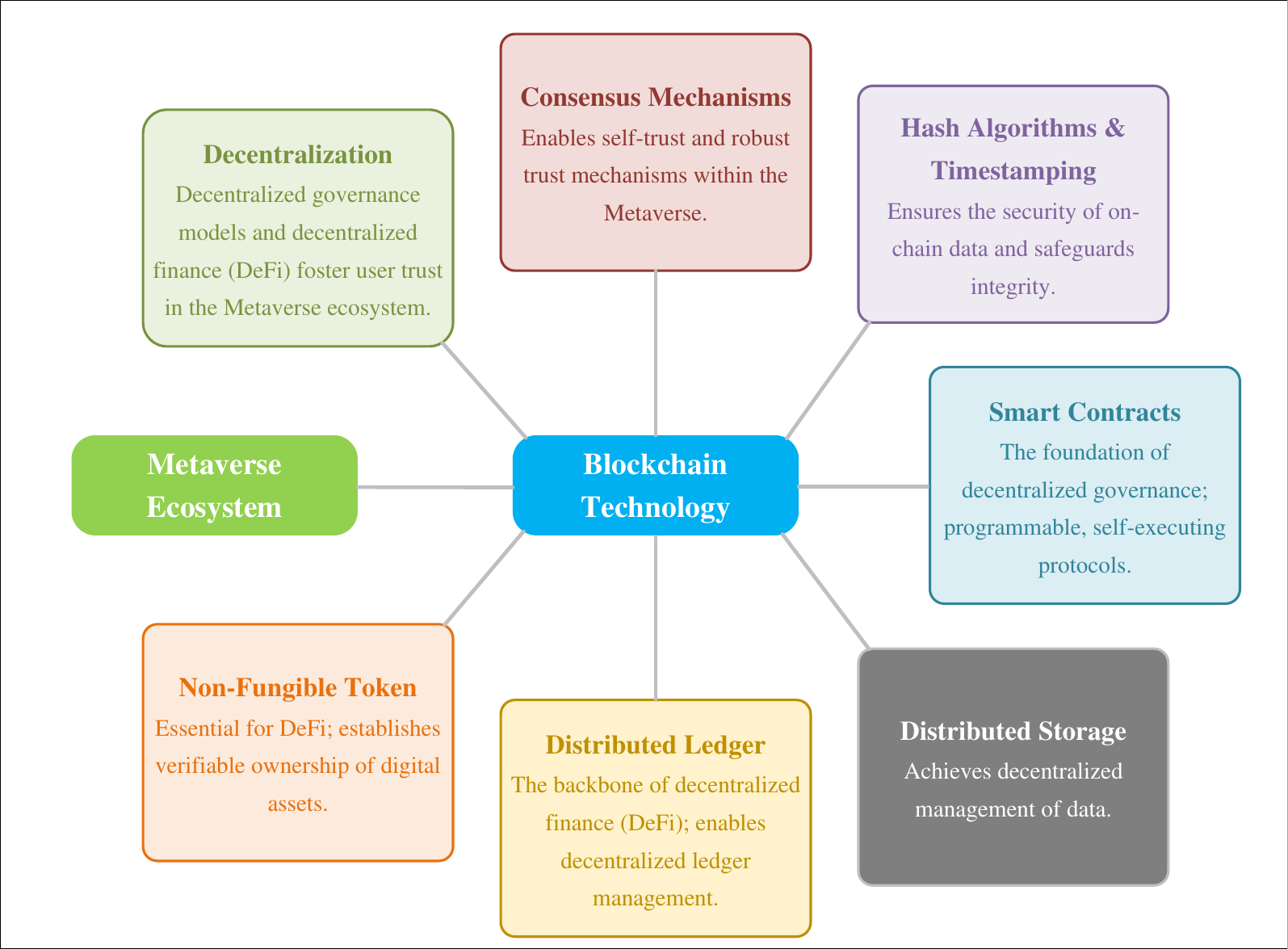}
  \caption{Paper Structure.}
  \label{fig:paper_structure}
  \vspace{-2ex}
\end{figure}

The organization of the paper is structured in Fig. \ref{fig:paper_structure}. Section \ref{sec::metaverse} provides a foundational overview of the Metaverse and blockchain, setting the stage for a deeper exploration of their interplay in subsequent sections. Section \ref{sec::decentralization} highlights decentralization as a cornerstone for addressing Metaverse challenges, while Section \ref{sec::consensus} details consensus mechanisms that establish trust in decentralized networks. Section \ref{sec::hash} introduces hash algorithms and timestamping, which ensure blockchain data security—prerequisites for reliable on-chain operations. Sections \ref{sec::smart_contract}, \ref{sec::distributed_storage}, \ref{sec::distributed_ledger} and \ref{sec::nft} explore critical blockchain applications in the Metaverse ecosystem: smart contracts (decentralized governance and automation), distributed storage (decentralized data management), distributed ledgers (decentralized finance and asset tracking), and NFTs (virtual-physical asset conversion). Finally, Section \ref{sec::conclusion} synthesizes the findings, emphasizing blockchain’s role in advancing the Metaverse through decentralization, security, and interoperability.

\section{BACKGROUND}
\label{sec::metaverse}

\subsection{Metaverse}
The concept of the Metaverse was first articulated in Neal Stephenson's seminal work, Snow Crash, wherein it is envisioned as a virtual reality environment \cite{mystakidis2022metaverse}. At the time of its conception, this notion was undoubtedly intriguing; however, the technological limitations of that era rendered the realization of such a Metaverse unattainable. In recent years, however, significant advancements in various technologies—most notably in 5G, Extended Reality (XR), Blockchain, and Artificial Intelligence—have catalyzed a renewed interest in the feasibility of Metaverses \cite{wang2022survey}. 

This shift has amplified public interest in the allure of the Metaverse and expanded its market, subsequently attracting substantial investment from private capital and corporations. Such dynamics have collectively propelled the evolution of the Metaverse into a prominent area of focus within technological discourse \cite{sparkes2021metaverse}. The integration of various technologies, along with their interplay with the real world, has given rise to a digital habitat characterized by innovative social systems—this is the contemporary understanding of the Metaverse.

\subsection{Blockchain}
Establishing a viable Metaverse is heavily contingent upon applying advanced technologies, with blockchain as a pivotal element \cite{nofer2017blockchain}. Blockchain, which originated with the advent of Bitcoin, is defined as a decentralized data structure that organizes information into blocks for verification and storage \cite{yang2022fusing} \cite{mourtzis2023blockchain}. Blockchain facilitates the generation and updating of the data within its framework by employing distributed node consensus algorithms \cite{nakamoto2008bitcoin}. The distinguishing features of blockchain—namely decentralization, transparency, independence, and security—provide a compelling foundation for numerous applications; Bitcoin remains the preeminent example of blockchain's successful implementation. 

In light of Bitcoin's increasing prominence, blockchain has garnered considerable academic and industrial interest, leading to swift advancements and extensive research endeavors \cite{pilkington2016blockchain} \cite{yaga2019blockchain}. Organizations from diverse sectors actively seek to leverage blockchain to enhance their competitive positioning, fostering deeper insights into its applications. The maturation of blockchain is thus imparting robust technical support to the Metaverse, effectively serving as a key that unlocks the potential of this digital frontier.

\section{DECENTRALIZATION}
\label{sec::decentralization}
\begin{figure}[htbp]
  \centering
  \includegraphics[width=0.8\textwidth, trim=20 20 20 20, clip]{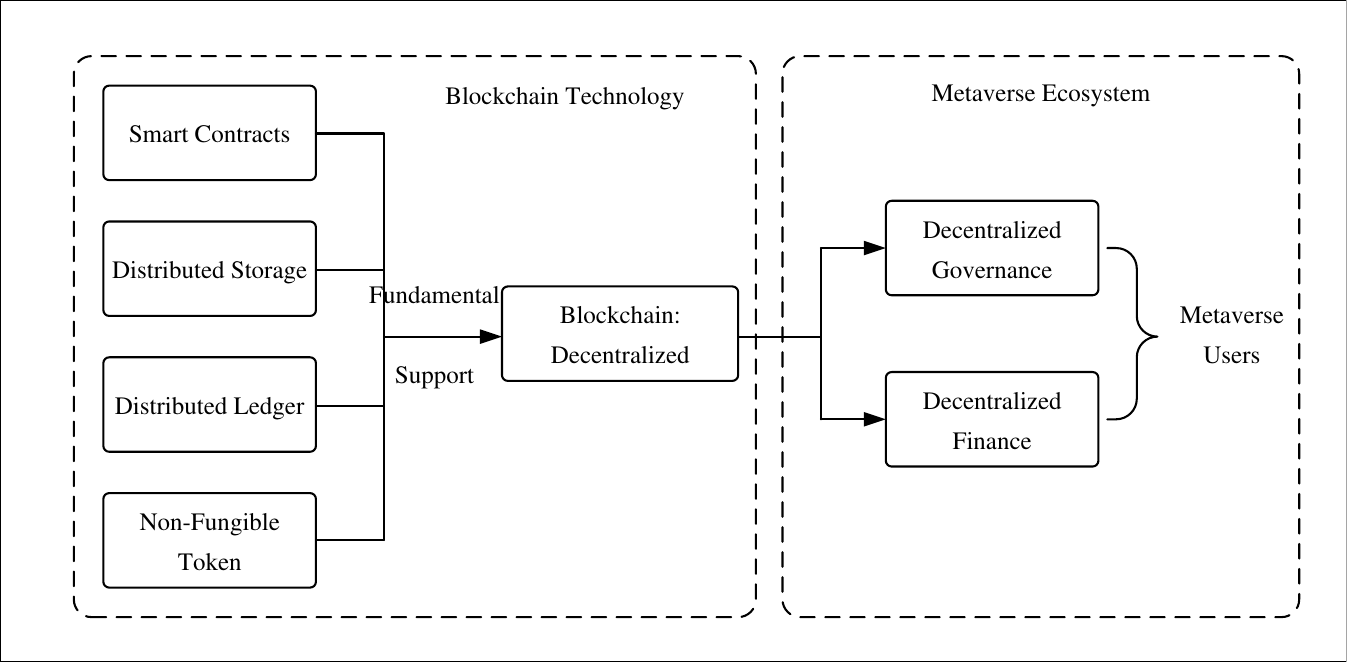}
  \caption{Decentralization to Blockchain and Metaverse Ecology.}
  \label{fig:relation}
  \vspace{-2ex}
\end{figure}

Decentralization is a key feature of blockchain and represents an alternative approach to centralized systems. In the blockchain Web 3.0 community, the term Metaverse has emerged to describe a decentralized virtual world \cite{ray2023web3}. The efforts towards establishing this decentralized existence had gained momentum as various companies explored the Metaverse, particularly in 2021 when significant interest from companies intensified. Many companies, driven by profit motives, aim to create centralized Metaverses that control user information and data. However, such centralized approaches may conflict with the community's desires for a truly decentralized experience \cite{dionisio20133d}. The ideal scenario for the Metaverse involves a decentralized, interconnected virtual environment supported by a decentralized governance model and decentralized finance, ensuring a user-centric experience. The relationship between decentralization, blockchain, and the Metaverse ecosystem is shown in Fig. \ref{fig:relation}.

\subsection{Decentralized Governance Model} 
Decentralization implies the absence of a central organization with ultimate decision-making authority within the Metaverse. A decentralized governance model is necessary to safeguard users' rights and interests while preventing monopoly influence to promote long-term stability in the Metaverse. This model represents a fundamental requirement from users regarding the Metaverse's ecology, with blockchain providing the optimal solution for achieving such governance. Blockchain facilitates the decentralization of the Metaverse, with the governance power residing with users. Deploying smart contracts and distributed storage will enable effective decentralized governance.

\subsection{Decentralized Finance} 
Decentralization extends beyond governance to include financial systems within the Metaverse \cite{li2024defitail}. As the Metaverse strives to integrate more closely with the real world, it aims to mitigate the boundaries between these realms, facilitating interactions between traditional and virtual financial systems \cite{zetzsche2020decentralized}. Trust in the safety of assets within this environment is essential for users. In conventional centralized economic systems, trust is often derived from national authority and regulatory frameworks. However, this may not be the case universally, affecting user confidence in state and corporate entities. Blockchain enables the formation of decentralized finance, which operates independently of traditional centralized structures. This new financial system fosters trust through user participation and coded protocols, promoting confidence in asset security within the Metaverse and supporting the dissolution of barriers between real and virtual economies.

The financial system is intrinsically linked to assets and asset flows. In a decentralized financial system, the management of these assets and flows must also be decentralized. Achieving this requires technical support from NFT, distributed ledgers, and smart contracts, all elements of blockchain. NFTs facilitate asset confirmation within the Metaverse ecosystem, while distributed ledgers provide real-time records accessible to all users. When combined with smart contracts, these technologies enable decentralized transactions and facilitate the flow of assets, forming the basis of a decentralized financial system in the Metaverse.

In contrast, the traditional centralized financial system relies on the trust associated with a central institution. Economic losses may occur if a central institution withdraws capital or experiences a data breach. Centralized institutions hold significant power but may lack a robust trust mechanism. The decentralized financial system, however, does not depend on a central authority but distributes rights among users. Asset identification, circulation data, and rights management are all decentralized. Users reach consensus through a consensus mechanism, while a hash algorithm and time-stamping safeguard data. This structure aims to establish a reliable trust mechanism within the decentralized finance ecosystem of the Metaverse \cite{liu2025sok}. 

The Metaverse, as a virtual world reflecting real-world economic activities, necessitates user trust, which can be achieved through decentralization in governance and finance. Confidence in the decentralized finance systems within the Metaverse and expectations for economic activities in this virtual space are increasing.

\subsection{Summary}
The Metaverse, closely tied to the real world, encompasses various user activities and must establish user trust by decentralizing its governance and financial models. Blockchain provides a framework for this decentralized approach, addressing questions related to its structure, governance, and financial systems. This foundation is essential for gaining user trust and establishing a governance model and economic system for the Metaverse. The application of decentralization through blockchain—utilizing smart contracts, distributed storage, distributed ledgers, and NFTs—aims to protect user rights and assets, reduce barriers between the real and virtual worlds, and advance the integration of the Metaverse into everyday life \cite{wang2023survey}.

\section{CONSENSUS MECHANISMS}
\label{sec::consensus}
Decentralization enables users to establish trust within the Metaverse ecosystem. In this decentralized environment, users must trust one another and reach consensus. Unlike a centralized network, where all nodes follow commands from a central node, the Metaverse operates without a central authority. Therefore, nodes must still function collaboratively and establish agreement to prevent disagreements. Within the Metaverse ecosystem, all nodes must maintain equal status without the emergence of centralized nodes while ensuring effective operation and consensus among all participants. The consensus mechanism provided by blockchain is positioned to fulfill the requirements for adaptation in the Metaverse ecosystem \cite{lashkari2021comprehensive}.

\subsection{Concepts} 
In blockchain, the consensus mechanism facilitates the verification and confirmation of transactions through a voting process among nodes, thus allowing nodes to trust one another and reach agreements. The consensus mechanism enables self-trust among blockchain nodes, eliminating the need for centralized institutions to ensure data consistency and validity \cite{zheng2017overview}. In the context of the Metaverse ecosystem, mutual trust among user nodes is a foundational requirement for decentralization. Furthermore, users within the ecosystem must establish self-trust, which can be achieved through agreements that function as the consensus mechanism.

\subsection{Application and Significance} 
The consensus mechanism serves as the foundational technology of blockchain, creating a network of trust among machines and codes and facilitating node trust within the ecosystem. Its application in the Metaverse promotes mutual trust among users and supports self-trust within the ecosystem \cite{gadekallu2022blockchain}. Additionally, it provides essential technical support for various applications, including decentralized governance, decentralized finance, smart contracts, NFTs, distributed ledgers, and distributed storage—core technologies of blockchain that depend on inter-node trust. The consensus mechanism is critical for enabling this mutual trust among nodes.

\begin{figure}[htbp]
  \centering
  \includegraphics[width=0.7\textwidth, trim=20 20 20 20, clip]{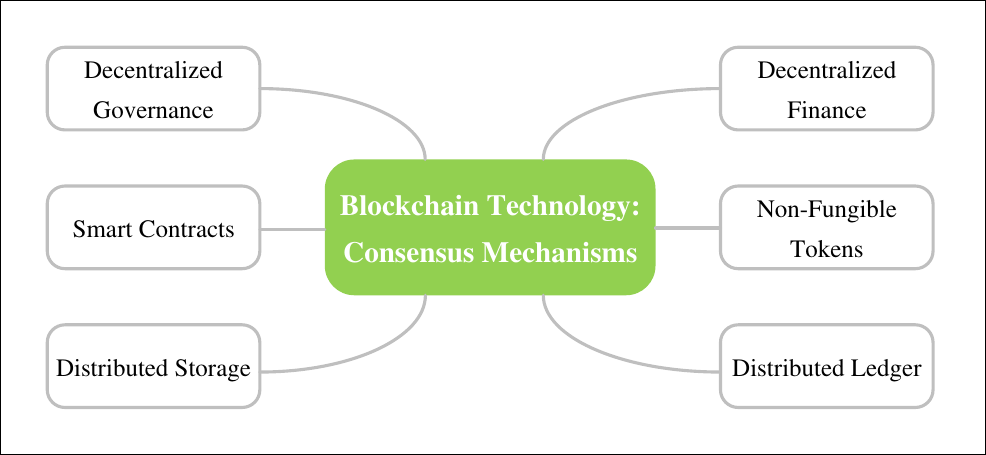}
  \caption{Enabling the Metaverse through Blockchain Underpinned by the Consensus Mechanism.}
  \label{fig:consensus_mechanism}
  \vspace{-2ex}
\end{figure}

The consensus mechanism enables self-trust within the Metaverse ecosystem while ensuring data consistency and validity. It allows users to reach agreements that align their interests and foster collaboration while ensuring that data among nodes remains consistent and valid. In blockchain, the consensus mechanism is fundamental to decentralization and is the foundation for node cohesion. Similarly, within the Metaverse ecosystem, it underpins mutual trust among user nodes, supports decentralization, and effectively maintains data consistency and validity in a decentralized environment. The application of the consensus mechanism is shown in Fig. \ref{fig:consensus_mechanism}.

\subsection{Mainstream Consensus Mechanisms}
Various consensus mechanisms have been developed to address the requirements of different application scenarios in the Metaverse ecosystem. The mainstream consensus mechanisms include proof-of-work (PoW), proof-of-stake (PoS), delegated proof-of-stake (DPoS), and practical Byzantine fault-tolerant algorithm (PBFT) \cite{castro1999practical}.

\begin{itemize}
  \item \textbf{PoW:} The proof-of-work mechanism relies on the computing power of distributed nodes that compete to solve a specific puzzle, SHA256. The first node to solve the puzzle earns the right to add a new block to the blockchain. PoW achieves decentralization through the computational efforts of nodes, though it is characterized by high resource consumption.
  \item \textbf{PoS:} The proof-of-stake mechanism utilizes cryptocurrencies that can be converted into coin age, which reflects network performance. This method of achieving consensus relies on the internal incentives from the coin age rather than computational power, resulting in reduced resource use and shorter processing times, albeit at the expense of some decentralization.
  \item \textbf{DPoS:} Delegated proof-of-stake involves nodes with voting rights electing a limited number of nodes for block generation. This approach reduces the number of nodes engaged in consensus verification, enhancing efficiency and reducing the degree of decentralization.
  \item \textbf{PBFT:} The Practical Byzantine Fault Tolerant algorithm allows a node to initiate a vote as long as fewer than (N-1)/3 of the nodes oppose it, where N represents the total number of nodes. PBFT is mathematically designed to handle up to (N-1)/3 faulty nodes. However, for the algorithm to function effectively, more than (N-1)/2 nodes must be operational.
\end{itemize}

\subsection{Summary}
Decentralization is a fundamental characteristic of the Metaverse ecosystem, serving as a basis for mutual trust among users. In a decentralized Metaverse, the absence of a central authority necessitates that users and nodes establish trust among themselves, which is facilitated by consensus mechanisms. These mechanisms enable users and nodes to reach agreements, promoting collaboration and aligning the interests of each participant while minimizing disputes. Consequently, consensus mechanisms are essential for achieving decentralization in the Metaverse and facilitating self-trust within the ecosystem \cite{xu2023trustless}. Additionally, there is a need for security algorithms to protect the Metaverse ecosystem further, thereby supporting the trust and self-trust of its users \cite{nguyen2022metachain}.

\section{HASH ALGORITHM AND TIMESTAMPING}
\label{sec::hash}
In blockchain, hash algorithms and timestamps generate blocks and confirm transaction integrity. Each block contains a target hash value, and mining involves miners calculating random solutions until they achieve this target hash. Upon successfully generating a block, a timestamp is assigned. All transaction data is converted into a hash value and stored within the blockchain. The unique characteristics of hash algorithms ensure that the data cannot be tampered with, thus maintaining transaction integrity and validity. In the Metaverse ecosystem, hash algorithms and timestamp are applied to ensure data integrity and confidentiality, while timestamps provide traceability for data.

\subsection{Hash Algorithm} 
A hash algorithm is a method that converts messages of arbitrary length into a fixed-length output, and it falls under the category of cryptography. These algorithms are characterized by their one-way function, which means that deducing the input value from the hash value is not feasible. They are also designed to exhibit weak and strong collision resistance: it should not be possible to find two different inputs that result in the same hash value, and discovering any pair of distinct inputs that produce the same hash value is infeasible. While the input can be arbitrary, the output hash value is consistently fixed in length \cite{wang2017survey}. 

The primary hash algorithm utilized in blockchain is HASH-256, which produces a 256-bit hash value typically represented in a 64-character hexadecimal format. The implementation process includes initializing constants, appending length values to the original message, and iterating through loops to derive the final hash value.

\subsection{Timestamping} 
\begin{figure}[htbp]
  \centering
  \includegraphics[width=0.7\textwidth, trim=20 20 20 20, clip]{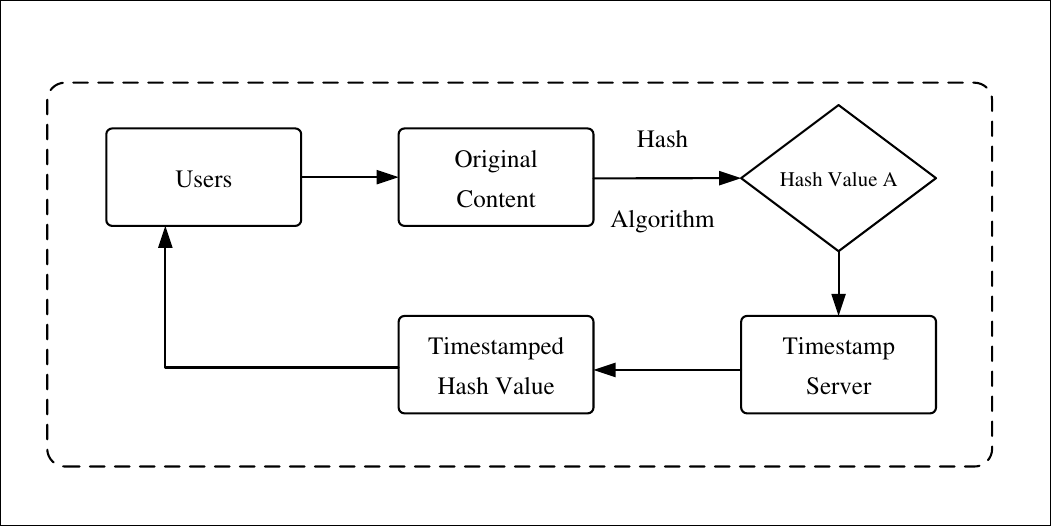}
  \caption{The Working Principle of Timestamping.}
  \label{fig:timestamp}
  \vspace{-2ex}
\end{figure}

Timestamping is a technical method used to assign a mark to a data segment, facilitating the verification of its integrity and establishing the time of existence \cite{denning1981timestamps}. The working principle of timestamp is shown in Fig. \ref{fig:timestamp}. A timestamp comprises three components: the file digest, the timestamp server's receipt time of the file, and the server's signature. 
The formation of a timestamp follows a defined process: the user applies a hash algorithm to the file, submits the processed file to a timestamp server for the timestamp application, and upon receipt, the server performs signature processing on both the hash value and the time record. The completed timestamp is then returned to the user, finalizing the timestamping process.

\subsection{Application}
The proposal and development of Metaverse ecology have led to the widespread application of blockchain in the Metaverse. Hash algorithms and timestamp play a significant role in ensuring the security of Metaverse ecology through blockchain \cite{zhang2019security}. These technologies are essential for blockchain functions such as block confirmation, the ordering of blocks, safeguarding data security, and verifying transaction integrity.

\begin{figure}[htbp]
  \centering
  \includegraphics[width=0.9\textwidth, trim=20 20 20 20, clip]{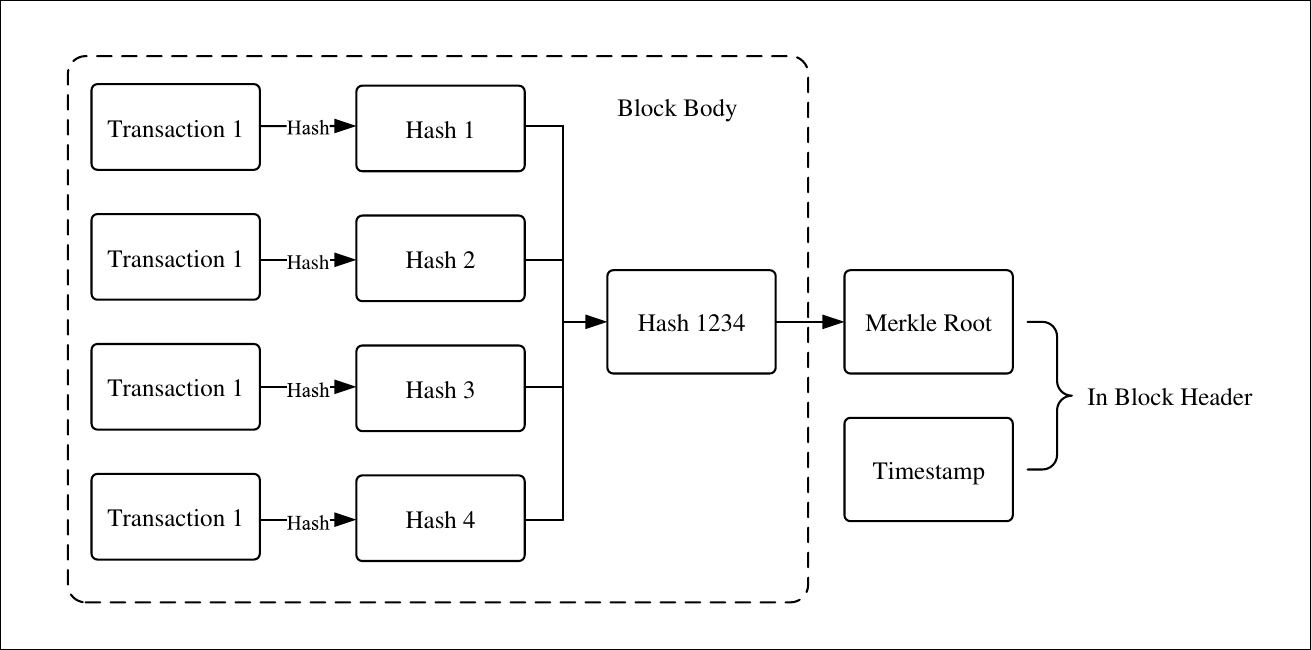}
  \caption{Hashing Algorithms and Timestamps in Blockchain.}
  \label{fig:hash_and_timestamp}
  \vspace{-2ex}
\end{figure}

Fig. \ref{fig:hash_and_timestamp} shows the hash algorithm and timestamp application in the blockchain. In the context of the Metaverse ecology, substantial amounts of data must be stored on the blockchain. The data is converted into hash values using hash algorithms, and a Merkle root is created based on a Merkle structure, which is then included in the block header alongside a timestamp before being added to the main chain. This process completes the uplinking operation of crucial data \cite{di2017blockchain}. By utilizing hash algorithms and timestamp, the security and integrity of the blockchain data can be maintained, ensuring that the uploaded data in the Metaverse remains secure and intact.

The application of hash algorithms and timestamp protects data security within the Metaverse ecosystem. However, the extensive data involved in the ecosystem can lead to inefficiencies and increased costs. Consequently, future considerations may involve limiting the data uploaded to the blockchain to only core information while employing alternative data storage and encryption methods.
Hash algorithms and timestamp applications have expanded to areas such as digital signatures and intellectual property protection. Future developments may further enhance the applications of these technologies within the Metaverse ecosystem.

\subsection{Summary}
Hash algorithms and timestamp are critical components of blockchain, providing security for data integrity. As blockchain continues to be integrated into the Metaverse ecosystem, these technologies will play a pivotal role in ensuring data security and integrity. Ongoing advancements in hash algorithms and timestamp suggest that they will perform additional functions and see increased applications within the Metaverse in the future.

\section{SMART CONTRACTS}
\label{sec::smart_contract}
The future of the Metaverse is anticipated to be decentralized, allowing users to participate in its management. Blockchain's central consensus mechanism enables self-trust among user nodes, while hash algorithms and timestamping technologies ensure data security. Smart contracts are the foundation for decentralized governance, facilitating user involvement in managing the Metaverse.

\subsection{Concepts}
The concept of smart contracts was introduced in 1995, but it gained significance only after the advent of Bitcoin. With the rise of blockchain, smart contracts have become integral to its functionality. A smart contract stores the rules and logic of a pre-agreed arrangement on the blockchain, executing its terms when specific conditions are met. Automation, programmability, tamper-proof features, and decentralization characterize this process \cite{szabo1996smart}. A common real-world example of a smart contract is the operation of vending machines.

\begin{figure}[htbp]
  \centering
  \includegraphics[width=0.9\textwidth, trim=20 20 20 20, clip]{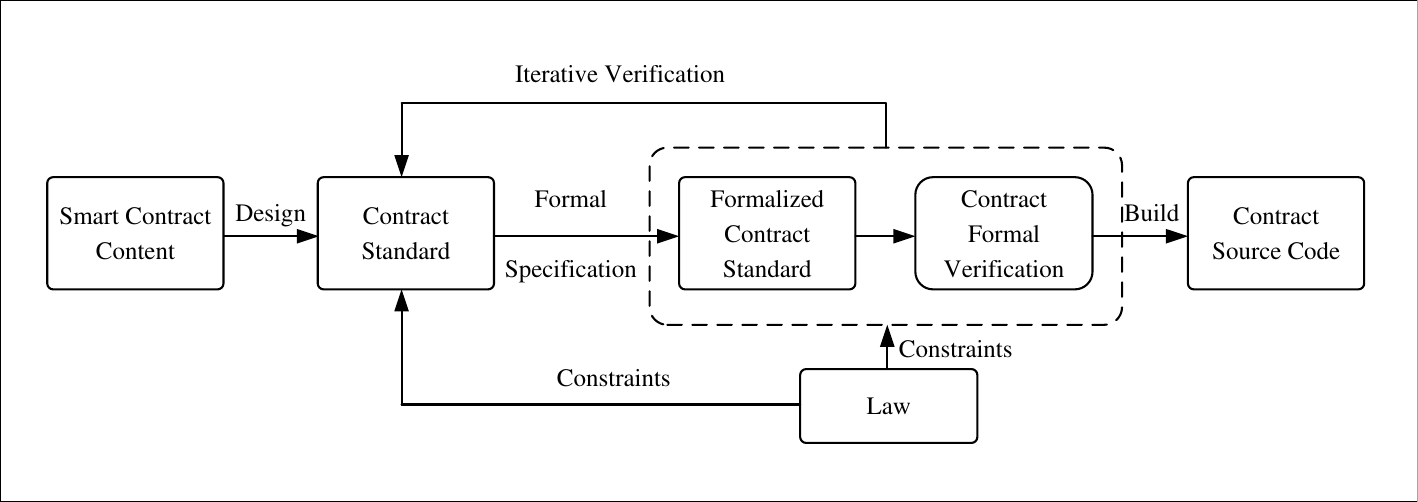}
  \caption{Smart Contract Creation Process.}
  \label{fig:contract_creation}
  \vspace{-2ex}
\end{figure}

Fig. \ref{fig:contract_creation} shows the smart contract creation process. Smart contracts operate by encoding the commitments of participating users into code. This code encompasses the required executions, states, scenarios, and actions that trigger the contract, accompanied by a description of the rules \cite{li2024scla} \cite{li2025scalm}. The source code is compiled into a system-required format \cite{li2024cobra} \cite{bu2025smartbugbert} (bytecode for the blockchain) before being integrated into transaction data. This transaction data is transmitted through a P2P network and stored on the blockchain. 
Each contract is assigned a unique contract address, which records and tracks its details. The smart contract verifies the state of its internal conditions and scenarios and then sends the eligible contract for validation \cite{li2021hybrid}. This contract propagates to verification nodes, which sign off upon successful verification. The execution of the contract begins once a consensus is reached.

\subsection{Application}
Decentralization is a primary requirement within the Metaverse, encompassing governance and financial systems that rely on blockchain, specifically smart contracts. These contracts can be utilized in various aspects of the Metaverse, including social interactions, organizational management, and even decentralized governance. Additionally, smart contracts play a significant role in decentralized finance, GameFi, the digital economy of the Metaverse, and trading platforms \cite{wang2024smart}. 

\begin{figure}[htbp]
  \centering
  \includegraphics[width=0.8\textwidth, trim=20 20 20 20, clip]{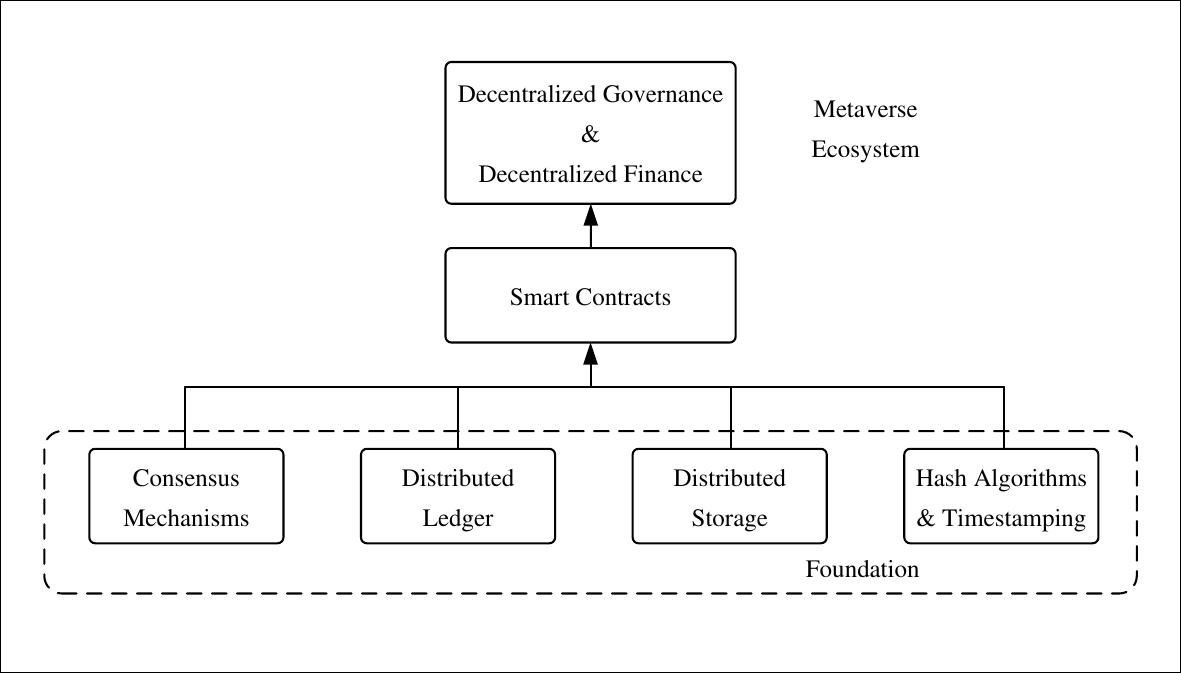}
  \caption{The Underlying Blockchain for Smart Contracts in Metaverse.}
  \label{fig:foundation}
  \vspace{-2ex}
\end{figure}

The role of smart contracts in the Metaverse ecosystem is shown in Fig. \ref{fig:foundation}. In a decentralized Metaverse ecosystem, no central authority allows smart contracts to manage the environment effectively. This framework aligns with user demands for decentralization, enabling all users to collaborate on contract content and execute agreements independently. Each user is assigned a unique identification, and multiple users can create a smart contract through mutual agreement. Signing the smart contract validates user participation and ensures the agreed terms are implemented. Once deployed on the blockchain, smart contracts are accessible for viewing and verification by all users, fostering trust in the contract's validity and facilitating decentralized governance and financial interactions without reliance on third-party intermediaries.

Smart contracts possess specific characteristics that contribute to user trust. Firstly, the content and rules of a smart contract are established through mutual agreement among users, ensuring recognition and acceptance by all parties involved. Once a smart contract is activated and entered into the blockchain, it is stored in a distributed manner, which prevents arbitrary tampering. The automatic execution of the contract's terms upon triggering predefined conditions minimizes human intervention and potential errors \cite{kolvart2016smart} . This inherent fairness and impartiality in execution lend credibility to smart contracts. 

Additionally, applied research continues to advance smart contracts, focusing on enhancing their efficiency and intelligence \cite{buterin2014next}. Smart contracts are primarily utilized in simple booking scenarios, often structured around IF-THEN logic. Future developments aim to adapt smart contracts to diverse scenarios and improve their formulation and execution processes \cite{khan2021blockchain}. Although they already demonstrate enhanced efficiency compared to traditional contracts, further improvements in storage efficiency and other areas are still possible.

However, smart contracts also face security risks. The visibility of distributed storage allows for the potential exposure of vulnerabilities that malicious actors could exploit. Addressing these vulnerabilities requires significant time and resources, highlighting the importance of developing multifaceted solutions, particularly within Metaverse applications, to mitigate potential losses caused by such security issues \cite{li2017discovering} \cite{bu2025enhancing}.

\subsection{Summary}
Smart contracts hold significant implications for the Metaverse, providing the foundational technical architecture for decentralized governance models and financial systems. They enable programmable and automated execution mechanisms, facilitating decentralization within the Metaverse ecosystem. 
Continued development and refinement of smart contracts are essential for their integration within the Metaverse. Initially adapted for blockchain, smart contracts have evolved into a foundational element that underpins the architecture of the Metaverse. They are central to enabling decentralization, serving as a core component of decentralized governance and finance. The implementation of smart contracts facilitates user participation in Metaverse management, contributing to the realization of decentralization.

\section{DISTRIBUTED STORAGE}
\label{sec::distributed_storage}
In the era of big data, data sensitivity and security have become increasingly prominent concerns, particularly within the context of the Metaverse ecosystem. For users to trust that their actions within this digital environment are genuine and effective, they must have control over their data. Distributed storage provides a framework for users to maintain ownership of their information. In a blockchain system, every node retains complete block data, and mechanisms such as consensus protocols, data transmission standards, and verification processes ensure the authenticity and security of this data. Thus, distributed storage not only facilitates decentralization within the Metaverse ecosystem but also lays the foundation for automation, helping to foster user trust.

\subsection{Distributed Storage in Blockchain}  
Distributed storage is a technique for managing data by distributing datasets across multiple locations \cite{benisi2020blockchain}. This approach is fundamental to blockchain, where each node stores the entirety of the blockchain's data. Key characteristics of distributed storage necessitate that data requirements remain consistent across different nodes to ensure data integrity. If any node fails to sustain, the system remains operational, and data continuity is maintained through network protocols.

Blockchain safeguards against tampering and ensures data verifiability, thus maintaining data consistency across nodes through hash and timestamps. The system's resilience allows it to continue functioning generally despite node failures, and recovery mechanisms enable the restoration of block data. Additionally, network disruptions do not impede the functioning of operational nodes, and disconnected nodes can reintegrate seamlessly once network connectivity is restored.

\subsection{Application}  
The Metaverse ecosystem is a digital representation closely connected to the physical world, resulting in substantial data requirements. Effective storage solutions are critical for developing the Metaverse to manage this extensive data landscape. Two primary types of storage are prevalent today: centralized storage and decentralized distributed storage \cite{khalid2023comprehensive}. While centralized storage offers a variety of devices with strong enhancements and mature market compatibility, its centralized nature can conflict with the Metaverse's demand for decentralization. This centralization can create dependencies that may become bottlenecks in data processing capacity.  

Conversely, decentralized distributed storage aligns more closely with the objectives of the Metaverse, featuring simple architecture, scalability, and suitability for automation. In this model, data is distributed across nodes, creating redundancy and improving security and efficiency. By promoting decentralization and automation, distributed storage is an essential factor in the sustainable and secure operation of the Metaverse ecosystem, assisting in the management of data flow within this digital environment.

Distributed storage aims to optimize the efficiency of storing and utilizing large volumes of data within the Metaverse ecosystem while maintaining data security. The requirements of the Metaverse include a robust distributed storage solution with guaranteed data security. Blockchain provides a suitable framework for distributed storage due to its characteristics of high redundancy, decentralization, and strong security. It ensures all block data is completely stored across each node, contributing to its permanence \cite{karaarslan2020data}. In this architecture, the authority of a central node is distributed among all nodes, enhancing efficiency through a peer-to-peer transmission protocol. This efficiency holds substantial commercial potential for the Metaverse ecosystem.

\subsection{Exploration of Distributed Storage}  
BitTorrent (BT) is an early solution for centralized storage, employing a peer-to-peer file transfer model that eliminates the need for central authority in data transmission. Users who receive files can then share them with others; however, the absence of an incentive mechanism can lead to the loss of files during transmission if not retained. The blockchain-based public chain Tiangui MEMO, developed by Chinese scientists, is pivotal in facilitating distributed storage within the Metaverse ecosystem. It establishes a significant technical architecture for both distributed storage and national strategic technological infrastructure \cite{memolabs2025}.

The Interplanetary File System (IPFS) protocol represents a prominent application of distributed storage. It integrates established concepts such as Distributed Hash Tables (DHT), Self-Authenticating File Systems (SFS), Merkle Directed Acyclic Graphs, Version Management Systems (Git), BT protocols, and blockchain to enhance file storage and content distribution \cite{benet2014ipfs}. DHT serves as a routing function to assist user nodes in locating data, while SFS addresses file naming for efficient transmission. The Merkle Directed Acyclic Graph enables content addressing, avoids data duplication, and protects against tampering. The Version Management System assists with file updates and historical version retrieval.

In comparison to HTTP, the IPFS protocol utilizes content addressing, accommodates greater decentralization, provides faster download speeds, and offers more economical storage solutions. Its capabilities include permanent file storage that resists tampering, secure encryption algorithms \cite{doan2022toward}, and an architecture composed of eight layers: the application layer, naming layer, file layer, object layer, switching layer, routing layer, network layer, and identity layer \cite{trautwein2022design}. As shown in Table \ref{tab:ipfs_stack}. The IPFS protocol enhances the coordination of file content across multiple users, supports permanent storage, and aims to improve efficiency while minimizing resource waste.

\begin{table}[ht]
  \small
  \centering
  \caption{Eight-Layer Protocol Stack of IPFS.}
  \vspace{-1ex}
  \label{tab:ipfs_stack}
  \begin{tabular}{>{\centering\arraybackslash}m{3cm} >{\raggedright\arraybackslash}m{10cm}} 
      \toprule
      \textbf{Layer} & \textbf{Description}\\
      \midrule
      Identity Layer	& Peer identity information generation.\\
      \midrule
      Network Layer	& Arbitrary transport layer protocols: ICE and NAT traversal.\\
      \midrule
      Routing Layer	& Distributed Sloppy Hash Table (DSHT) locates peers and stores required object information.\\
      \midrule
      Exchange Layer	& Manages block distribution.\\
      \midrule
      Object Layer	& Merkle-DAG: Content-addressable, tamper-proof, deduplicated object links.\\
      \midrule
      File Layer	& Git-like versioned file system.\\
      \midrule
      Naming Layer & Assigns and modifies names for objects.\\
      \midrule
      Application Layer	& Applications running on IPFS utilize nearest-node service provision to improve efficiency and reduce costs.\\
      \bottomrule
  \end{tabular}
  \end{table}

\subsection{Summary}
In the context of advancing toward the Metaverse, distributed storage serves as a fundamental technology within the Metaverse ecosystem. Numerous technology enterprises are actively exploring distributed storage, although the practical applications currently in use are limited and face certain technical challenges. For instance, the scale of nodes remains relatively constrained, and service quality lacks established standards and guarantees. Additionally, there is a need to improve the incentive layer, specifically regarding data valuation and the balance of incentive mechanisms. The protocol layer requires optimization of code to enhance data reading and writing capabilities and to improve overall efficiency \cite{dimakis2006decentralized}. 

As a foundational technology in the Metaverse ecosystem, distributed storage has significant potential for enhancement, ensuring it can effectively support the extensive data requirements of the Metaverse and provide a solid foundation for its development.
Distributed storage supports decentralization in the Metaverse ecosystem, enabling true decentralization in data storage. It is also integral to blockchain, as distributed storage facilitates smart contracts, distributed ledgers, and NFTs, making it essential within the Metaverse framework.

\section{DISTRIBUTED LEDGER}
\label{sec::distributed_ledger}
Bitcoin often exemplifies Distributed ledger, which operates on blockchain principles. Bitcoin represents a digital currency that relies on a distributed ledger. In this system, a miner node engages in mining operations to obtain the bookkeeping rights for a block, which the system rewards \cite{sunyaev2020distributed}. Once the node secures these rights, it stores Bitcoin transactions for a specified duration in a new block, links the block to the blockchain, and broadcasts the information to other nodes. Upon receiving the broadcast, these nodes verify and store the data, resulting in the cumulative recording of Bitcoin transactions within a distributed ledger. This established application illustrates the characteristics and advantages of distributed ledgers.

\subsection{Concepts}
\begin{figure}[htbp]
  \centering
  \includegraphics[width=0.6\textwidth, trim=20 20 20 20, clip]{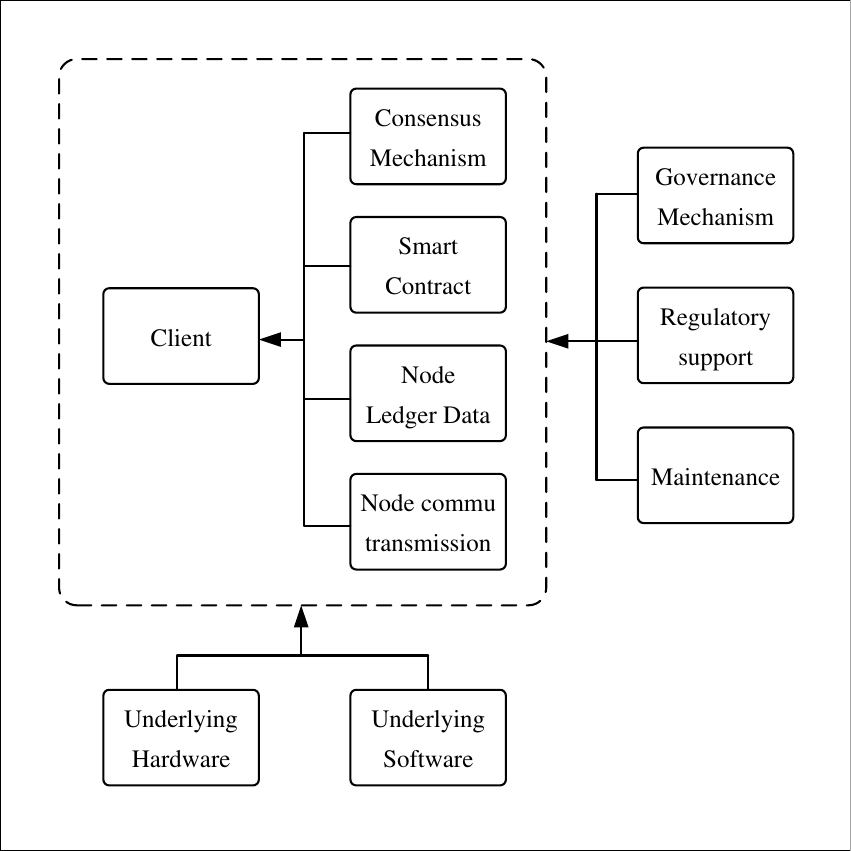}
  \caption{Distributed Ledger Infrastructure.}
  \label{fig:distributed_ledger}
  \vspace{-2ex}
\end{figure}

The basic architecture of the distributed ledger is shown in Fig. \ref{fig:distributed_ledger}. A distributed ledger functions as a shared, replicated, and synchronized database among network participants. Its distributed nature is closely related to distributed storage. In contrast, distributed storage focuses primarily on content storage, and distributed ledgers are concerned with recording transactions among member nodes and maintaining an auditable historical record of all transactions within the network.

As an innovative database form, a distributed ledger ensures that all member nodes possess a copy of the ledger, with any changes reflected across all copies. This process occurs rapidly in a peer-to-peer network, promoting swift communication and data transfer. Decentralization, tamper-proof attributes, and ease of traceability characterize the distributed ledger \cite{li2024guardians}.

\subsection{Application}

In a completely decentralized financial system within the Metaverse ecosystem, the assets generated and contributed by users play a critical role in facilitating the flow of financial activity. A ledger that ensures the legitimacy, traceability, and validity of transactions is essential to this ecosystem. Given the Metaverse's decentralized nature, the ledger must also be decentralized, accessible to all users, and efficient in operation \cite{el2018review}.

The requirements for this ledger within the Metaverse ecosystem suggest that a distributed ledger offers a suitable solution. Distributed ledgers feature storage across multiple nodes, ensuring decentralization and tamper-resistance, as each node retains all transaction records. All transaction data must be validated before being uploaded and broadcasted, ensuring that the information recorded on the blockchain is legal and valid \cite{liu2024gastrace}. 

Additionally, due to the inherent qualities of blockchain, data stored on the distributed ledger has a permanent existence, facilitating the secure retention of all transactional records for future reference \cite{li2024detecting} \cite{li2024stateguard}. Peer-to-peer transmission of bookkeeping data allows efficient traversal across all nodes, thus enhancing bookkeeping efficiency and improving reconciliation processes. This design also addresses transaction trust issues, increasing transactional efficiency and minimizing resource wastage resulting from trust-related challenges \cite{rauchs2018distributed}.

The application of distributed ledgers is particularly prominent in finance, aligning well with the financial systems of the Metaverse. As a core technology, distributed ledgers will provide foundational ledger services for the financial system within this ecosystem. In addition to hardware support, the efficacy of distributed ledgers depends on various underlying blockchain technologies, including consensus mechanisms, smart contracts, node ledger data, and node communication protocols, collectively enhancing security. The governance mechanism oversees nodes' organizational structures and interactions with the distributed ledger, while regulatory support establishes protocols and technologies to optimize ledger design and functionality \cite{maull2017distributed}. Regular maintenance and updates of the distributed ledger contribute to its overall security.

Distributed ledgers demonstrate advantages such as decentralization, peer-to-peer transmission, data integrity, and traceability through a combination of hardware architecture and underlying software. These qualities contribute to establishing a reliable, efficient ledger system for financial operations within the Metaverse, forming the foundation for transaction platforms and trust mechanisms in this ecological environment.

\subsection{Summary}
The distributed ledger, which serves as the foundational blockchain, supports the requirements of a decentralized Metaverse ecosystem. It enables a decentralized financial system within the Metaverse, enhances transaction efficiency, reduces costs, and emerges as a central technology for this environment. Distributed ledger has found numerous applications and continues to evolve, significantly focusing on integrating with the Metaverse ecology. This integration is essential because the Metaverse serves as a key application for distributed ledgers. It enables users to convert real assets into Metaverse assets. As distributed ledger evolves within this ecosystem, it also expands its applications and functionalities in the Metaverse \cite{natarajan2017distributed}.

\section{NON-FUNGIBLE TOKENS}
\label{sec::nft}
The Metaverse gives users confidence in its ecosystem through decentralization, a feature further supported by a decentralized financial infrastructure. This arrangement assures users of the financial system's safety, encouraging them to invest assets within the Metaverse, which subsequently necessitates a means to verify asset rights within the environment.

\subsection{Concepts}
Non-fungible tokens (NFTs) are a type of blockchain-based asset distinguished from fungible tokens such as Bitcoin. Fungible tokens represent uniform currencies where each unit is interchangeable, while non-fungible tokens maintain uniqueness and irreplaceability. This distinction signifies that the properties of NFTs—uniqueness, traceability, and indivisibility—provide significant advantages for asset verification \cite{wang2021non} \cite{razi2023non}.

\begin{figure}[htbp]
  \centering
  \includegraphics[width=0.8\textwidth, trim=20 20 20 20, clip]{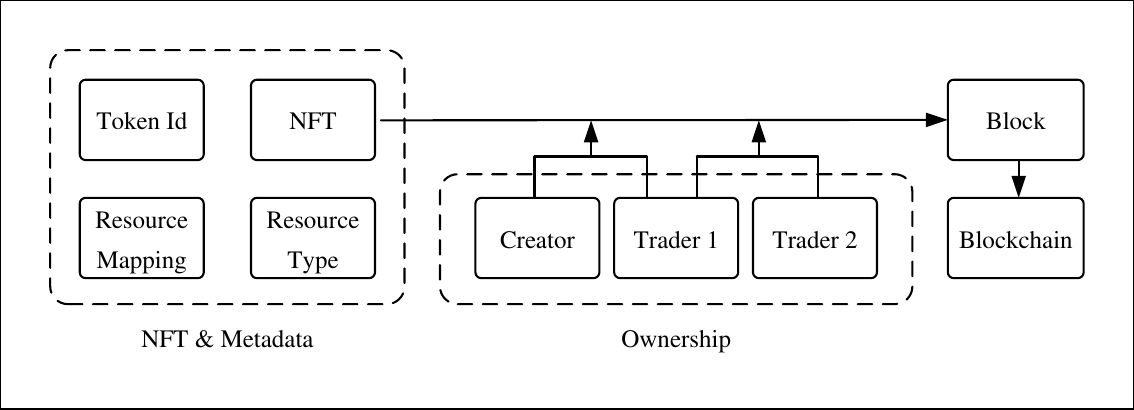}
  \caption{The Working Logic for NFTs.}
  \label{fig:nft}
  \vspace{-2ex}
\end{figure}

The functionality of NFTs is realized through a process that begins by assigning a unique identifier to the item intended for tokenization. This identifier must represent the item and ensure its uniqueness. The identifier is then tokenized and placed on a public blockchain chain. The general process includes selecting a suitable blockchain public chain, developing smart contracts on that chain while adhering to established standards, and then deploying these contracts to create a decentralized application (DApp). The identification data and contract rules are stored on the blockchain, thus creating the NFT and facilitating the storage of ownership information that can be displayed and modified by the smart contract. The working principle of NFT is shown in Fig. \ref{fig:nft}.

\subsection{Application}

The Metaverse facilitates decentralized finance, enabling users to engage with the financial systems within the Metaverse with a level of trust. By integrating their assets into this ecosystem, users can earn and circulate assets, contributing to the active nature of the financial system. In contrast to the traditional financial landscape, where asset ownership is validated through currency issued by centralized institutions, the Metaverse operates without such central authorities. It raises the question of how individuals can prove their ownership of assets in a decentralized environment.

In the Metaverse, all assets are digital, and there is a significant demand for decentralized methods to validate ownership. Non-fungible tokens (NFTs) play a crucial role in establishing ownership of assets within the Metaverse ecosystem. Decentralized finance ensures that real-world assets are incorporated into the Metaverse. At the same time, NFTs validate ownership rights, confirming that the asset is uniquely attributed to the user who injected it and recognized throughout the Metaverse ecology. NFTs confirm the uniqueness of asset ownership and establish recognition within the entire ecosystem \cite{taherdoost2022non}.

Once ownership is confirmed, economic circulation can occur within the Metaverse. As assets change hands, NFTs can update ownership information, and all transaction details can be recorded on a public blockchain. This process enhances asset safety during circulation and allows for traceability to the asset's origin. The operational capabilities of NFTs regarding asset authentication and traceability address trust issues that can arise during transactions in the Metaverse. Information stored on the public chain gains consensus recognition among Metaverse users, establishing a transparent framework where asset ownership is visible and verifiable by all parties.

The presence of NFTs can alleviate trust problems between transaction parties, thereby improving transaction efficiency and reducing resource waste associated with trust issues. Additionally, NFTs authenticate asset ownership and safeguard copyrights within the Metaverse \cite{kong2024characterizing}. Users can create digital objects and utilize NFTs to assert and protect the rights to their creations. 

Beyond asset authentication and copyright protection, the functionalities of NFTs can extend to identity verification within the Metaverse. Users' various identities can be minted into NFTs and recorded on the public chain, providing a basis for verifying the authenticity of user identities. This application of NFTs facilitates the operation of identity validation in a decentralized manner, simplifying processes within the Metaverse ecosystem overall \cite{das2022understanding}.

\subsection{Future Directions}

NFTs are characterized by their uniqueness, indivisibility, non-replicability, and traceability. They are widely used for the authentication of virtual assets and the digital certification of physical assets. On some occasions, NFTs are primarily traded as a form of cryptocurrency. These places have adopted more liberal policies regarding public blockchains and secondary markets, resulting in a greater availability of NFTs. It has led to a supply that often exceeds demand, with many high-priced NFTs frequently being traded and speculation occurring in the market \cite{niu2024unveiling}.

On other occasions, NFTs are primarily viewed as digital collectibles. Regulatory approaches to the public blockchain are more stringent, prompting the use of alliance chains—networks formed by specific groups—with lower degrees of decentralization. This results in reduced circulation of NFTs and heightened regulation of the secondary market to mitigate speculative risks and financial concerns. Consequently, pricing tends to be cheaper and more uniform. Regardless of future developments, the trajectory indicates a direction toward integration within the Metaverse ecosystem, where NFTs serve a core role.

\subsection{Summary}

NFTs function as blockchain tokens with distinct features such as uniqueness, indivisibility, traceability, and non-replicability, serving as critical infrastructure within the Metaverse ecosystem \cite{dong2023copyright}. Within this context, NFTs facilitate asset verification and copyright protection, primarily focusing on asset authentication. This capability underpins decentralized finance in the Metaverse, contributing to integrating real and virtual economies, enhancing the efficiency of financial transactions, and fostering a more streamlined financial system.

\section{CONCLUSION}
\label{sec::conclusion}

This study highlights the central role of blockchain in building the Metaverse's infrastructure, focusing on key components such as decentralization, consensus mechanisms, hash algorithms, smart contracts, distributed storage, ledgers, and NFTs. Together, these technologies create a secure, transparent, and decentralized foundation.
Decentralization, a core tenet of blockchain, addresses governance and financial requirements within the Metaverse. Consensus mechanisms foster trust among participants in a decentralized environment, while smart contracts enable automated governance and expanded functionality. Hash algorithms and timestamping enhance security, whereas distributed storage and ledgers optimize data integrity and transactional transparency. NFTs enrich the Metaverse economy by enabling verifiable asset ownership without centralized intermediaries.
Though marked by rapid growth, the Metaverse is still in its early stages. Blockchain’s flexibility and alignment with its core requirements position it as a vital enabler. As both technologies evolve, blockchain will drive future innovations and bridge virtual and physical worlds.

\bibliographystyle{ACM-Reference-Format}
\bibliography{acmart}


\begin{thebibliography}{61}


\ifx \showCODEN    \undefined \def \showCODEN     #1{\unskip}     \fi
\ifx \showDOI      \undefined \def \showDOI       #1{#1}\fi
\ifx \showISBNx    \undefined \def \showISBNx     #1{\unskip}     \fi
\ifx \showISBNxiii \undefined \def \showISBNxiii  #1{\unskip}     \fi
\ifx \showISSN     \undefined \def \showISSN      #1{\unskip}     \fi
\ifx \showLCCN     \undefined \def \showLCCN      #1{\unskip}     \fi
\ifx \shownote     \undefined \def \shownote      #1{#1}          \fi
\ifx \showarticletitle \undefined \def \showarticletitle #1{#1}   \fi
\ifx \showURL      \undefined \def \showURL       {\relax}        \fi
\providecommand\bibfield[2]{#2}
\providecommand\bibinfo[2]{#2}
\providecommand\natexlab[1]{#1}
\providecommand\showeprint[2][]{arXiv:#2}

\bibitem[Benet(2014)]%
        {benet2014ipfs}
\bibfield{author}{\bibinfo{person}{Juan Benet}.} \bibinfo{year}{2014}\natexlab{}.
\newblock \showarticletitle{Ipfs-content addressed, versioned, p2p file system}.
\newblock \bibinfo{journal}{\emph{arXiv preprint arXiv:1407.3561}} (\bibinfo{year}{2014}).
\newblock


\bibitem[Benisi et~al\mbox{.}(2020)]%
        {benisi2020blockchain}
\bibfield{author}{\bibinfo{person}{Nazanin~Zahed Benisi}, \bibinfo{person}{Mehdi Aminian}, {and} \bibinfo{person}{Bahman Javadi}.} \bibinfo{year}{2020}\natexlab{}.
\newblock \showarticletitle{Blockchain-based decentralized storage networks: A survey}.
\newblock \bibinfo{journal}{\emph{Journal of Network and Computer Applications}}  \bibinfo{volume}{162} (\bibinfo{year}{2020}), \bibinfo{pages}{102656}.
\newblock


\bibitem[Bu et~al\mbox{.}(2025a)]%
        {bu2025enhancing}
\bibfield{author}{\bibinfo{person}{Jiuyang Bu}, \bibinfo{person}{Wenkai Li}, \bibinfo{person}{Zongwei Li}, \bibinfo{person}{Zeng Zhang}, {and} \bibinfo{person}{Xiaoqi Li}.} \bibinfo{year}{2025}\natexlab{a}.
\newblock \showarticletitle{Enhancing Smart Contract Vulnerability Detection in DApps Leveraging Fine-Tuned LLM}.
\newblock \bibinfo{journal}{\emph{arXiv preprint arXiv:2504.05006}} (\bibinfo{year}{2025}).
\newblock


\bibitem[Bu et~al\mbox{.}(2025b)]%
        {bu2025smartbugbert}
\bibfield{author}{\bibinfo{person}{Jiuyang Bu}, \bibinfo{person}{Wenkai Li}, \bibinfo{person}{Zongwei Li}, \bibinfo{person}{Zeng Zhang}, {and} \bibinfo{person}{Xiaoqi Li}.} \bibinfo{year}{2025}\natexlab{b}.
\newblock \showarticletitle{SmartBugBert: BERT-Enhanced Vulnerability Detection for Smart Contract Bytecode}.
\newblock \bibinfo{journal}{\emph{arXiv preprint arXiv:2504.05002}} (\bibinfo{year}{2025}).
\newblock


\bibitem[Buterin et~al\mbox{.}(2014)]%
        {buterin2014next}
\bibfield{author}{\bibinfo{person}{Vitalik Buterin} {et~al\mbox{.}}} \bibinfo{year}{2014}\natexlab{}.
\newblock \showarticletitle{A next-generation smart contract and decentralized application platform}.
\newblock \bibinfo{journal}{\emph{white paper}} \bibinfo{volume}{3}, \bibinfo{number}{37} (\bibinfo{year}{2014}), \bibinfo{pages}{2--1}.
\newblock


\bibitem[Castro et~al\mbox{.}(1999)]%
        {castro1999practical}
\bibfield{author}{\bibinfo{person}{Miguel Castro}, \bibinfo{person}{Barbara Liskov}, {et~al\mbox{.}}} \bibinfo{year}{1999}\natexlab{}.
\newblock \showarticletitle{Practical byzantine fault tolerance}. In \bibinfo{booktitle}{\emph{OsDI}}, Vol.~\bibinfo{volume}{99}. \bibinfo{pages}{173--186}.
\newblock


\bibitem[Das et~al\mbox{.}(2022)]%
        {das2022understanding}
\bibfield{author}{\bibinfo{person}{Dipanjan Das}, \bibinfo{person}{Priyanka Bose}, \bibinfo{person}{Nicola Ruaro}, \bibinfo{person}{Christopher Kruegel}, {and} \bibinfo{person}{Giovanni Vigna}.} \bibinfo{year}{2022}\natexlab{}.
\newblock \showarticletitle{Understanding security issues in the NFT ecosystem}. In \bibinfo{booktitle}{\emph{Proceedings of the 2022 ACM SIGSAC Conference on Computer and Communications Security}}. \bibinfo{pages}{667--681}.
\newblock


\bibitem[Denning and Sacco(1981)]%
        {denning1981timestamps}
\bibfield{author}{\bibinfo{person}{Dorothy~E Denning} {and} \bibinfo{person}{Giovanni~Maria Sacco}.} \bibinfo{year}{1981}\natexlab{}.
\newblock \showarticletitle{Timestamps in key distribution protocols}.
\newblock \bibinfo{journal}{\emph{Commun. ACM}} \bibinfo{volume}{24}, \bibinfo{number}{8} (\bibinfo{year}{1981}), \bibinfo{pages}{533--536}.
\newblock


\bibitem[Di~Pierro(2017)]%
        {di2017blockchain}
\bibfield{author}{\bibinfo{person}{Massimo Di~Pierro}.} \bibinfo{year}{2017}\natexlab{}.
\newblock \showarticletitle{What is the blockchain?}
\newblock \bibinfo{journal}{\emph{Computing in Science \& Engineering}} \bibinfo{volume}{19}, \bibinfo{number}{5} (\bibinfo{year}{2017}), \bibinfo{pages}{92--95}.
\newblock


\bibitem[Dimakis et~al\mbox{.}(2006)]%
        {dimakis2006decentralized}
\bibfield{author}{\bibinfo{person}{Alexandros~G Dimakis}, \bibinfo{person}{Vinod Prabhakaran}, {and} \bibinfo{person}{Kannan Ramchandran}.} \bibinfo{year}{2006}\natexlab{}.
\newblock \showarticletitle{Decentralized erasure codes for distributed networked storage}.
\newblock \bibinfo{journal}{\emph{IEEE Transactions on Information Theory}} \bibinfo{volume}{52}, \bibinfo{number}{6} (\bibinfo{year}{2006}), \bibinfo{pages}{2809--2816}.
\newblock


\bibitem[Dionisio et~al\mbox{.}(2013)]%
        {dionisio20133d}
\bibfield{author}{\bibinfo{person}{John David~N Dionisio}, \bibinfo{person}{William G~Burns Iii}, {and} \bibinfo{person}{Richard Gilbert}.} \bibinfo{year}{2013}\natexlab{}.
\newblock \showarticletitle{3D virtual worlds and the metaverse: Current status and future possibilities}.
\newblock \bibinfo{journal}{\emph{ACM computing surveys (CSUR)}} \bibinfo{volume}{45}, \bibinfo{number}{3} (\bibinfo{year}{2013}), \bibinfo{pages}{1--38}.
\newblock


\bibitem[Doan et~al\mbox{.}(2022)]%
        {doan2022toward}
\bibfield{author}{\bibinfo{person}{Trinh~Viet Doan}, \bibinfo{person}{Yiannis Psaras}, \bibinfo{person}{J{\"o}rg Ott}, {and} \bibinfo{person}{Vaibhav Bajpai}.} \bibinfo{year}{2022}\natexlab{}.
\newblock \showarticletitle{Toward decentralized cloud storage with IPFS: opportunities, challenges, and future considerations}.
\newblock \bibinfo{journal}{\emph{IEEE Internet Computing}} \bibinfo{volume}{26}, \bibinfo{number}{6} (\bibinfo{year}{2022}), \bibinfo{pages}{7--15}.
\newblock


\bibitem[Dong and Wang(2023)]%
        {dong2023copyright}
\bibfield{author}{\bibinfo{person}{Yupeng Dong} {and} \bibinfo{person}{Chunhui Wang}.} \bibinfo{year}{2023}\natexlab{}.
\newblock \showarticletitle{Copyright protection on NFT digital works in the Metaverse}.
\newblock \bibinfo{journal}{\emph{Security and Safety}}  \bibinfo{volume}{2} (\bibinfo{year}{2023}), \bibinfo{pages}{2023013}.
\newblock


\bibitem[El~Ioini and Pahl(2018)]%
        {el2018review}
\bibfield{author}{\bibinfo{person}{Nabil El~Ioini} {and} \bibinfo{person}{Claus Pahl}.} \bibinfo{year}{2018}\natexlab{}.
\newblock \showarticletitle{A review of distributed ledger technologies}. In \bibinfo{booktitle}{\emph{On the Move to Meaningful Internet Systems. OTM 2018 Conferences: Confederated International Conferences: CoopIS, C\&TC, and ODBASE 2018, Valletta, Malta, October 22-26, 2018, Proceedings, Part II}}. Springer, \bibinfo{pages}{277--288}.
\newblock


\bibitem[Gadekallu et~al\mbox{.}(2022)]%
        {gadekallu2022blockchain}
\bibfield{author}{\bibinfo{person}{Thippa~Reddy Gadekallu}, \bibinfo{person}{Thien Huynh-The}, \bibinfo{person}{Weizheng Wang}, \bibinfo{person}{Gokul Yenduri}, \bibinfo{person}{Pasika Ranaweera}, \bibinfo{person}{Quoc-Viet Pham}, \bibinfo{person}{Daniel~Benevides da Costa}, {and} \bibinfo{person}{Madhusanka Liyanage}.} \bibinfo{year}{2022}\natexlab{}.
\newblock \showarticletitle{Blockchain for the metaverse: A review}.
\newblock \bibinfo{journal}{\emph{arXiv preprint arXiv:2203.09738}} (\bibinfo{year}{2022}).
\newblock


\bibitem[Karaarslan and Konacakl{\i}(2020)]%
        {karaarslan2020data}
\bibfield{author}{\bibinfo{person}{Enis Karaarslan} {and} \bibinfo{person}{Enis Konacakl{\i}}.} \bibinfo{year}{2020}\natexlab{}.
\newblock \showarticletitle{Data storage in the decentralized world: Blockchain and derivatives}.
\newblock \bibinfo{journal}{\emph{arXiv preprint arXiv:2012.10253}} (\bibinfo{year}{2020}).
\newblock


\bibitem[Khalid et~al\mbox{.}(2023)]%
        {khalid2023comprehensive}
\bibfield{author}{\bibinfo{person}{Muhammad~Irfan Khalid}, \bibinfo{person}{Ibtisam Ehsan}, \bibinfo{person}{Ayman~Khallel Al-Ani}, \bibinfo{person}{Jawaid Iqbal}, \bibinfo{person}{Saddam Hussain}, \bibinfo{person}{Syed~Sajid Ullah}, {et~al\mbox{.}}} \bibinfo{year}{2023}\natexlab{}.
\newblock \showarticletitle{A comprehensive survey on blockchain-based decentralized storage networks}.
\newblock \bibinfo{journal}{\emph{IEEE Access}}  \bibinfo{volume}{11} (\bibinfo{year}{2023}), \bibinfo{pages}{10995--11015}.
\newblock


\bibitem[Khan et~al\mbox{.}(2021)]%
        {khan2021blockchain}
\bibfield{author}{\bibinfo{person}{Shafaq~Naheed Khan}, \bibinfo{person}{Faiza Loukil}, \bibinfo{person}{Chirine Ghedira-Guegan}, \bibinfo{person}{Elhadj Benkhelifa}, {and} \bibinfo{person}{Anoud Bani-Hani}.} \bibinfo{year}{2021}\natexlab{}.
\newblock \showarticletitle{Blockchain smart contracts: Applications, challenges, and future trends}.
\newblock \bibinfo{journal}{\emph{Peer-to-peer Networking and Applications}}  \bibinfo{volume}{14} (\bibinfo{year}{2021}), \bibinfo{pages}{2901--2925}.
\newblock


\bibitem[Kolvart et~al\mbox{.}(2016)]%
        {kolvart2016smart}
\bibfield{author}{\bibinfo{person}{Merit Kolvart}, \bibinfo{person}{Margus Poola}, {and} \bibinfo{person}{Addi Rull}.} \bibinfo{year}{2016}\natexlab{}.
\newblock \showarticletitle{Smart contracts}.
\newblock \bibinfo{journal}{\emph{The Future of Law and etechnologies}} (\bibinfo{year}{2016}), \bibinfo{pages}{133--147}.
\newblock


\bibitem[Kong et~al\mbox{.}(2024)]%
        {kong2024characterizing}
\bibfield{author}{\bibinfo{person}{Dechao Kong}, \bibinfo{person}{Xiaoqi Li}, {and} \bibinfo{person}{Wenkai Li}.} \bibinfo{year}{2024}\natexlab{}.
\newblock \showarticletitle{Characterizing the Solana NFT ecosystem}. In \bibinfo{booktitle}{\emph{Companion Proceedings of the ACM Web Conference 2024}}. \bibinfo{pages}{766--769}.
\newblock


\bibitem[Labs(2025)]%
        {memolabs2025}
\bibfield{author}{\bibinfo{person}{Memo Labs}.} \bibinfo{year}{2025}\natexlab{}.
\newblock \bibinfo{title}{AI-Driven User Centric Data Modular Blockchain}.
\newblock \bibinfo{howpublished}{Website}.
\newblock
\newblock
\shownote{\url{https://memolabs.org/}}.


\bibitem[Lashkari and Musilek(2021)]%
        {lashkari2021comprehensive}
\bibfield{author}{\bibinfo{person}{Bahareh Lashkari} {and} \bibinfo{person}{Petr Musilek}.} \bibinfo{year}{2021}\natexlab{}.
\newblock \showarticletitle{A comprehensive review of blockchain consensus mechanisms}.
\newblock \bibinfo{journal}{\emph{IEEE access}}  \bibinfo{volume}{9} (\bibinfo{year}{2021}), \bibinfo{pages}{43620--43652}.
\newblock


\bibitem[Li et~al\mbox{.}(2024a)]%
        {li2024cobra}
\bibfield{author}{\bibinfo{person}{Wenkai Li}, \bibinfo{person}{Xiaoqi Li}, \bibinfo{person}{Zongwei Li}, {and} \bibinfo{person}{Yuqing Zhang}.} \bibinfo{year}{2024}\natexlab{a}.
\newblock \showarticletitle{Cobra: interaction-aware bytecode-level vulnerability detector for smart contracts}. In \bibinfo{booktitle}{\emph{Proceedings of the 39th IEEE/ACM International Conference on Automated Software Engineering}}. \bibinfo{pages}{1358--1369}.
\newblock


\bibitem[Li et~al\mbox{.}(2024d)]%
        {li2024defitail}
\bibfield{author}{\bibinfo{person}{Wenkai Li}, \bibinfo{person}{Xiaoqi Li}, \bibinfo{person}{Yuqing Zhang}, {and} \bibinfo{person}{Zongwei Li}.} \bibinfo{year}{2024}\natexlab{d}.
\newblock \showarticletitle{DeFiTail: DeFi Protocol Inspection through Cross-Contract Execution Analysis}. In \bibinfo{booktitle}{\emph{Companion Proceedings of the ACM Web Conference 2024}}. \bibinfo{pages}{786--789}.
\newblock


\bibitem[Li et~al\mbox{.}(2024e)]%
        {li2024detecting}
\bibfield{author}{\bibinfo{person}{Wenkai Li}, \bibinfo{person}{Zhijie Liu}, \bibinfo{person}{Xiaoqi Li}, {and} \bibinfo{person}{Sen Nie}.} \bibinfo{year}{2024}\natexlab{e}.
\newblock \showarticletitle{Detecting Malicious Accounts in Web3 through Transaction Graph}. In \bibinfo{booktitle}{\emph{Proceedings of the 39th IEEE/ACM International Conference on Automated Software Engineering}}. \bibinfo{pages}{2482--2483}.
\newblock


\bibitem[Li et~al\mbox{.}(2021)]%
        {li2021hybrid}
\bibfield{author}{\bibinfo{person}{Xiaoqi Li} {et~al\mbox{.}}} \bibinfo{year}{2021}\natexlab{}.
\newblock \showarticletitle{Hybrid analysis of smart contracts and malicious behaviors in ethereum}.
\newblock \bibinfo{journal}{\emph{Hong Kong Polytechnic University}} (\bibinfo{year}{2021}).
\newblock


\bibitem[Li et~al\mbox{.}(2024f)]%
        {li2024scla}
\bibfield{author}{\bibinfo{person}{Xiaoqi Li}, \bibinfo{person}{Yingjie Mao}, \bibinfo{person}{Zexin Lu}, \bibinfo{person}{Wenkai Li}, {and} \bibinfo{person}{Zongwei Li}.} \bibinfo{year}{2024}\natexlab{f}.
\newblock \showarticletitle{SCLA: Automated Smart Contract Summarization via LLMs and Control Flow Prompt}.
\newblock \bibinfo{journal}{\emph{arXiv e-prints}} (\bibinfo{year}{2024}), \bibinfo{pages}{arXiv--2402}.
\newblock


\bibitem[Li et~al\mbox{.}(2017)]%
        {li2017discovering}
\bibfield{author}{\bibinfo{person}{Xiaoqi Li}, \bibinfo{person}{L Yu}, {and} \bibinfo{person}{XP Luo}.} \bibinfo{year}{2017}\natexlab{}.
\newblock \showarticletitle{On Discovering Vulnerabilities in Android Applications}.
\newblock In \bibinfo{booktitle}{\emph{Mobile Security and Privacy}}. \bibinfo{publisher}{Elsevier}, \bibinfo{pages}{155--166}.
\newblock


\bibitem[Li et~al\mbox{.}(2024b)]%
        {li2024guardians}
\bibfield{author}{\bibinfo{person}{Zongwei Li}, \bibinfo{person}{Wenkai Li}, \bibinfo{person}{Xiaoqi Li}, {and} \bibinfo{person}{Yuqing Zhang}.} \bibinfo{year}{2024}\natexlab{b}.
\newblock \showarticletitle{Guardians of the ledger: Protecting decentralized exchanges from state derailment defects}.
\newblock \bibinfo{journal}{\emph{IEEE Transactions on Reliability}} (\bibinfo{year}{2024}).
\newblock


\bibitem[Li et~al\mbox{.}(2024c)]%
        {li2024stateguard}
\bibfield{author}{\bibinfo{person}{Zongwei Li}, \bibinfo{person}{Wenkai Li}, \bibinfo{person}{Xiaoqi Li}, {and} \bibinfo{person}{Yuqing Zhang}.} \bibinfo{year}{2024}\natexlab{c}.
\newblock \showarticletitle{StateGuard: Detecting State Derailment Defects in Decentralized Exchange Smart Contract}. In \bibinfo{booktitle}{\emph{Companion Proceedings of the ACM Web Conference 2024}}. \bibinfo{pages}{810--813}.
\newblock


\bibitem[Li et~al\mbox{.}(2025)]%
        {li2025scalm}
\bibfield{author}{\bibinfo{person}{Zongwei Li}, \bibinfo{person}{Xiaoqi Li}, \bibinfo{person}{Wenkai Li}, {and} \bibinfo{person}{Xin Wang}.} \bibinfo{year}{2025}\natexlab{}.
\newblock \showarticletitle{SCALM: Detecting Bad Practices in Smart Contracts Through LLMs}.
\newblock \bibinfo{journal}{\emph{arXiv preprint arXiv:2502.04347}} (\bibinfo{year}{2025}).
\newblock


\bibitem[Liu and Li(2025)]%
        {liu2025sok}
\bibfield{author}{\bibinfo{person}{Zekai Liu} {and} \bibinfo{person}{Xiaoqi Li}.} \bibinfo{year}{2025}\natexlab{}.
\newblock \showarticletitle{SoK: Security Analysis of Blockchain-based Cryptocurrency}.
\newblock \bibinfo{journal}{\emph{arXiv preprint arXiv:2503.22156}} (\bibinfo{year}{2025}).
\newblock


\bibitem[Liu et~al\mbox{.}(2024)]%
        {liu2024gastrace}
\bibfield{author}{\bibinfo{person}{Zekai Liu}, \bibinfo{person}{Xiaoqi Li}, \bibinfo{person}{Hongli Peng}, {and} \bibinfo{person}{Wenkai Li}.} \bibinfo{year}{2024}\natexlab{}.
\newblock \showarticletitle{GasTrace: Detecting Sandwich Attack Malicious Accounts in Ethereum}. In \bibinfo{booktitle}{\emph{2024 IEEE International Conference on Web Services (ICWS)}}. IEEE, \bibinfo{pages}{1409--1411}.
\newblock


\bibitem[Maull et~al\mbox{.}(2017)]%
        {maull2017distributed}
\bibfield{author}{\bibinfo{person}{Roger Maull}, \bibinfo{person}{Phil Godsiff}, \bibinfo{person}{Catherine Mulligan}, \bibinfo{person}{Alan Brown}, {and} \bibinfo{person}{Beth Kewell}.} \bibinfo{year}{2017}\natexlab{}.
\newblock \showarticletitle{Distributed ledger technology: Applications and implications}.
\newblock \bibinfo{journal}{\emph{Strategic Change}} \bibinfo{volume}{26}, \bibinfo{number}{5} (\bibinfo{year}{2017}), \bibinfo{pages}{481--489}.
\newblock


\bibitem[Mourtzis et~al\mbox{.}(2023)]%
        {mourtzis2023blockchain}
\bibfield{author}{\bibinfo{person}{Dimitris Mourtzis}, \bibinfo{person}{John Angelopoulos}, {and} \bibinfo{person}{Nikos Panopoulos}.} \bibinfo{year}{2023}\natexlab{}.
\newblock \showarticletitle{Blockchain integration in the era of industrial metaverse}.
\newblock \bibinfo{journal}{\emph{Applied Sciences}} \bibinfo{volume}{13}, \bibinfo{number}{3} (\bibinfo{year}{2023}), \bibinfo{pages}{1353}.
\newblock


\bibitem[Mystakidis(2022)]%
        {mystakidis2022metaverse}
\bibfield{author}{\bibinfo{person}{Stylianos Mystakidis}.} \bibinfo{year}{2022}\natexlab{}.
\newblock \showarticletitle{Metaverse}.
\newblock \bibinfo{journal}{\emph{Encyclopedia}} \bibinfo{volume}{2}, \bibinfo{number}{1} (\bibinfo{year}{2022}), \bibinfo{pages}{486--497}.
\newblock


\bibitem[Nakamoto(2008)]%
        {nakamoto2008bitcoin}
\bibfield{author}{\bibinfo{person}{Satoshi Nakamoto}.} \bibinfo{year}{2008}\natexlab{}.
\newblock \showarticletitle{Bitcoin: A peer-to-peer electronic cash system}.
\newblock  (\bibinfo{year}{2008}).
\newblock


\bibitem[Natarajan et~al\mbox{.}(2017)]%
        {natarajan2017distributed}
\bibfield{author}{\bibinfo{person}{Harish Natarajan}, \bibinfo{person}{Solvej Krause}, {and} \bibinfo{person}{Helen Gradstein}.} \bibinfo{year}{2017}\natexlab{}.
\newblock \showarticletitle{Distributed ledger technology and blockchain}.
\newblock  (\bibinfo{year}{2017}).
\newblock


\bibitem[Nguyen et~al\mbox{.}(2022)]%
        {nguyen2022metachain}
\bibfield{author}{\bibinfo{person}{Cong~T Nguyen}, \bibinfo{person}{Dinh~Thai Hoang}, \bibinfo{person}{Diep~N Nguyen}, {and} \bibinfo{person}{Eryk Dutkiewicz}.} \bibinfo{year}{2022}\natexlab{}.
\newblock \showarticletitle{Metachain: A novel blockchain-based framework for metaverse applications}. In \bibinfo{booktitle}{\emph{2022 IEEE 95th Vehicular Technology Conference:(VTC2022-Spring)}}. IEEE, \bibinfo{pages}{1--5}.
\newblock


\bibitem[Niu et~al\mbox{.}(2024)]%
        {niu2024unveiling}
\bibfield{author}{\bibinfo{person}{Yuanzheng Niu}, \bibinfo{person}{Xiaoqi Li}, \bibinfo{person}{Hongli Peng}, {and} \bibinfo{person}{Wenkai Li}.} \bibinfo{year}{2024}\natexlab{}.
\newblock \showarticletitle{Unveiling wash trading in popular NFT markets}. In \bibinfo{booktitle}{\emph{Companion Proceedings of the ACM Web Conference 2024}}. \bibinfo{pages}{730--733}.
\newblock


\bibitem[Nofer et~al\mbox{.}(2017)]%
        {nofer2017blockchain}
\bibfield{author}{\bibinfo{person}{Michael Nofer}, \bibinfo{person}{Peter Gomber}, \bibinfo{person}{Oliver Hinz}, {and} \bibinfo{person}{Dirk Schiereck}.} \bibinfo{year}{2017}\natexlab{}.
\newblock \showarticletitle{Blockchain}.
\newblock \bibinfo{journal}{\emph{Business \& information systems engineering}}  \bibinfo{volume}{59} (\bibinfo{year}{2017}), \bibinfo{pages}{183--187}.
\newblock


\bibitem[Pilkington(2016)]%
        {pilkington2016blockchain}
\bibfield{author}{\bibinfo{person}{Marc Pilkington}.} \bibinfo{year}{2016}\natexlab{}.
\newblock \showarticletitle{Blockchain technology: principles and applications}.
\newblock In \bibinfo{booktitle}{\emph{Research handbook on digital transformations}}. \bibinfo{publisher}{Edward Elgar Publishing}, \bibinfo{pages}{225--253}.
\newblock


\bibitem[Rauchs et~al\mbox{.}(2018)]%
        {rauchs2018distributed}
\bibfield{author}{\bibinfo{person}{Michel Rauchs}, \bibinfo{person}{Andrew Glidden}, \bibinfo{person}{Brian Gordon}, \bibinfo{person}{Gina~C Pieters}, \bibinfo{person}{Martino Recanatini}, \bibinfo{person}{Fran{\c{c}}ois Rostand}, \bibinfo{person}{Kathryn Vagneur}, {and} \bibinfo{person}{Bryan~Zheng Zhang}.} \bibinfo{year}{2018}\natexlab{}.
\newblock \showarticletitle{Distributed ledger technology systems: A conceptual framework}.
\newblock \bibinfo{journal}{\emph{Available at SSRN 3230013}} (\bibinfo{year}{2018}).
\newblock


\bibitem[Ray(2023)]%
        {ray2023web3}
\bibfield{author}{\bibinfo{person}{Partha~Pratim Ray}.} \bibinfo{year}{2023}\natexlab{}.
\newblock \showarticletitle{Web3: A comprehensive review on background, technologies, applications, zero-trust architectures, challenges and future directions}.
\newblock \bibinfo{journal}{\emph{Internet of Things and Cyber-Physical Systems}}  \bibinfo{volume}{3} (\bibinfo{year}{2023}), \bibinfo{pages}{213--248}.
\newblock


\bibitem[Razi et~al\mbox{.}(2023)]%
        {razi2023non}
\bibfield{author}{\bibinfo{person}{Qaiser Razi}, \bibinfo{person}{Aryan Devrani}, \bibinfo{person}{Harshal Abhyankar}, \bibinfo{person}{G~Sai~Sesha Chalapathi}, \bibinfo{person}{Vikas Hassija}, {and} \bibinfo{person}{Mohsen Guizani}.} \bibinfo{year}{2023}\natexlab{}.
\newblock \showarticletitle{Non-fungible tokens (NFTs)—Survey of current applications, evolution, and future directions}.
\newblock \bibinfo{journal}{\emph{IEEE Open Journal of the Communications Society}}  \bibinfo{volume}{5} (\bibinfo{year}{2023}), \bibinfo{pages}{2765--2791}.
\newblock


\bibitem[Sparkes(2021)]%
        {sparkes2021metaverse}
\bibfield{author}{\bibinfo{person}{Matthew Sparkes}.} \bibinfo{year}{2021}\natexlab{}.
\newblock \bibinfo{title}{What is a metaverse}.
\newblock
\newblock


\bibitem[Sunyaev and Sunyaev(2020)]%
        {sunyaev2020distributed}
\bibfield{author}{\bibinfo{person}{Ali Sunyaev} {and} \bibinfo{person}{Ali Sunyaev}.} \bibinfo{year}{2020}\natexlab{}.
\newblock \showarticletitle{Distributed ledger technology}.
\newblock \bibinfo{journal}{\emph{Internet computing: Principles of distributed systems and emerging internet-based technologies}} (\bibinfo{year}{2020}), \bibinfo{pages}{265--299}.
\newblock


\bibitem[Szabo(1996)]%
        {szabo1996smart}
\bibfield{author}{\bibinfo{person}{Nick Szabo}.} \bibinfo{year}{1996}\natexlab{}.
\newblock \showarticletitle{Smart contracts: building blocks for digital markets}.
\newblock \bibinfo{journal}{\emph{EXTROPY: The Journal of Transhumanist Thought,(16)}} \bibinfo{volume}{18}, \bibinfo{number}{2} (\bibinfo{year}{1996}), \bibinfo{pages}{28}.
\newblock


\bibitem[Taherdoost(2022)]%
        {taherdoost2022non}
\bibfield{author}{\bibinfo{person}{Hamed Taherdoost}.} \bibinfo{year}{2022}\natexlab{}.
\newblock \showarticletitle{Non-fungible tokens (NFT): A systematic review}.
\newblock \bibinfo{journal}{\emph{Information}} \bibinfo{volume}{14}, \bibinfo{number}{1} (\bibinfo{year}{2022}), \bibinfo{pages}{26}.
\newblock


\bibitem[Trautwein et~al\mbox{.}(2022)]%
        {trautwein2022design}
\bibfield{author}{\bibinfo{person}{Dennis Trautwein}, \bibinfo{person}{Aravindh Raman}, \bibinfo{person}{Gareth Tyson}, \bibinfo{person}{Ignacio Castro}, \bibinfo{person}{Will Scott}, \bibinfo{person}{Moritz Schubotz}, \bibinfo{person}{Bela Gipp}, {and} \bibinfo{person}{Yiannis Psaras}.} \bibinfo{year}{2022}\natexlab{}.
\newblock \showarticletitle{Design and evaluation of IPFS: a storage layer for the decentralized web}. In \bibinfo{booktitle}{\emph{Proceedings of the ACM SIGCOMM 2022 Conference}}. \bibinfo{pages}{739--752}.
\newblock


\bibitem[Wang et~al\mbox{.}(2023)]%
        {wang2023survey}
\bibfield{author}{\bibinfo{person}{Hang Wang}, \bibinfo{person}{Huansheng Ning}, \bibinfo{person}{Yujia Lin}, \bibinfo{person}{Wenxi Wang}, \bibinfo{person}{Sahraoui Dhelim}, \bibinfo{person}{Fadi Farha}, \bibinfo{person}{Jianguo Ding}, {and} \bibinfo{person}{Mahmoud Daneshmand}.} \bibinfo{year}{2023}\natexlab{}.
\newblock \showarticletitle{A survey on the metaverse: The state-of-the-art, technologies, applications, and challenges}.
\newblock \bibinfo{journal}{\emph{IEEE Internet of Things Journal}} \bibinfo{volume}{10}, \bibinfo{number}{16} (\bibinfo{year}{2023}), \bibinfo{pages}{14671--14688}.
\newblock


\bibitem[Wang et~al\mbox{.}(2017)]%
        {wang2017survey}
\bibfield{author}{\bibinfo{person}{Jingdong Wang}, \bibinfo{person}{Ting Zhang}, \bibinfo{person}{Nicu Sebe}, \bibinfo{person}{Heng~Tao Shen}, {et~al\mbox{.}}} \bibinfo{year}{2017}\natexlab{}.
\newblock \showarticletitle{A survey on learning to hash}.
\newblock \bibinfo{journal}{\emph{IEEE transactions on pattern analysis and machine intelligence}} \bibinfo{volume}{40}, \bibinfo{number}{4} (\bibinfo{year}{2017}), \bibinfo{pages}{769--790}.
\newblock


\bibitem[Wang et~al\mbox{.}(2021)]%
        {wang2021non}
\bibfield{author}{\bibinfo{person}{Qin Wang}, \bibinfo{person}{Rujia Li}, \bibinfo{person}{Qi Wang}, {and} \bibinfo{person}{Shiping Chen}.} \bibinfo{year}{2021}\natexlab{}.
\newblock \showarticletitle{Non-fungible token (NFT): Overview, evaluation, opportunities and challenges}.
\newblock \bibinfo{journal}{\emph{arXiv preprint arXiv:2105.07447}} (\bibinfo{year}{2021}).
\newblock


\bibitem[Wang et~al\mbox{.}(2024)]%
        {wang2024smart}
\bibfield{author}{\bibinfo{person}{Yishun Wang}, \bibinfo{person}{Xiaoqi Li}, \bibinfo{person}{Shipeng Ye}, \bibinfo{person}{Lei Xie}, {and} \bibinfo{person}{Ju Xing}.} \bibinfo{year}{2024}\natexlab{}.
\newblock \showarticletitle{Smart contracts in the real world: A statistical exploration of external data dependencies}.
\newblock \bibinfo{journal}{\emph{arXiv preprint arXiv:2406.13253}} (\bibinfo{year}{2024}).
\newblock


\bibitem[Wang et~al\mbox{.}(2022)]%
        {wang2022survey}
\bibfield{author}{\bibinfo{person}{Yuntao Wang}, \bibinfo{person}{Zhou Su}, \bibinfo{person}{Ning Zhang}, \bibinfo{person}{Rui Xing}, \bibinfo{person}{Dongxiao Liu}, \bibinfo{person}{Tom~H Luan}, {and} \bibinfo{person}{Xuemin Shen}.} \bibinfo{year}{2022}\natexlab{}.
\newblock \showarticletitle{A survey on metaverse: Fundamentals, security, and privacy}.
\newblock \bibinfo{journal}{\emph{IEEE communications surveys \& tutorials}} \bibinfo{volume}{25}, \bibinfo{number}{1} (\bibinfo{year}{2022}), \bibinfo{pages}{319--352}.
\newblock


\bibitem[Xu et~al\mbox{.}(2023)]%
        {xu2023trustless}
\bibfield{author}{\bibinfo{person}{Minghui Xu}, \bibinfo{person}{Yihao Guo}, \bibinfo{person}{Qin Hu}, \bibinfo{person}{Zehui Xiong}, \bibinfo{person}{Dongxiao Yu}, {and} \bibinfo{person}{Xiuzhen Cheng}.} \bibinfo{year}{2023}\natexlab{}.
\newblock \showarticletitle{A trustless architecture of blockchain-enabled metaverse}.
\newblock \bibinfo{journal}{\emph{High-confidence computing}} \bibinfo{volume}{3}, \bibinfo{number}{1} (\bibinfo{year}{2023}), \bibinfo{pages}{100088}.
\newblock


\bibitem[Yaga et~al\mbox{.}(2019)]%
        {yaga2019blockchain}
\bibfield{author}{\bibinfo{person}{Dylan Yaga}, \bibinfo{person}{Peter Mell}, \bibinfo{person}{Nik Roby}, {and} \bibinfo{person}{Karen Scarfone}.} \bibinfo{year}{2019}\natexlab{}.
\newblock \showarticletitle{Blockchain technology overview}.
\newblock \bibinfo{journal}{\emph{arXiv preprint arXiv:1906.11078}} (\bibinfo{year}{2019}).
\newblock


\bibitem[Yang et~al\mbox{.}(2022)]%
        {yang2022fusing}
\bibfield{author}{\bibinfo{person}{Qinglin Yang}, \bibinfo{person}{Yetong Zhao}, \bibinfo{person}{Huawei Huang}, \bibinfo{person}{Zehui Xiong}, \bibinfo{person}{Jiawen Kang}, {and} \bibinfo{person}{Zibin Zheng}.} \bibinfo{year}{2022}\natexlab{}.
\newblock \showarticletitle{Fusing blockchain and AI with metaverse: A survey}.
\newblock \bibinfo{journal}{\emph{IEEE Open Journal of the Computer Society}}  \bibinfo{volume}{3} (\bibinfo{year}{2022}), \bibinfo{pages}{122--136}.
\newblock


\bibitem[Zetzsche et~al\mbox{.}(2020)]%
        {zetzsche2020decentralized}
\bibfield{author}{\bibinfo{person}{Dirk~A Zetzsche}, \bibinfo{person}{Douglas~W Arner}, {and} \bibinfo{person}{Ross~P Buckley}.} \bibinfo{year}{2020}\natexlab{}.
\newblock \showarticletitle{Decentralized finance}.
\newblock \bibinfo{journal}{\emph{Journal of Financial Regulation}} \bibinfo{volume}{6}, \bibinfo{number}{2} (\bibinfo{year}{2020}), \bibinfo{pages}{172--203}.
\newblock


\bibitem[Zhang et~al\mbox{.}(2019)]%
        {zhang2019security}
\bibfield{author}{\bibinfo{person}{Rui Zhang}, \bibinfo{person}{Rui Xue}, {and} \bibinfo{person}{Ling Liu}.} \bibinfo{year}{2019}\natexlab{}.
\newblock \showarticletitle{Security and privacy on blockchain}.
\newblock \bibinfo{journal}{\emph{ACM Computing Surveys (CSUR)}} \bibinfo{volume}{52}, \bibinfo{number}{3} (\bibinfo{year}{2019}), \bibinfo{pages}{1--34}.
\newblock


\bibitem[Zheng et~al\mbox{.}(2017)]%
        {zheng2017overview}
\bibfield{author}{\bibinfo{person}{Zibin Zheng}, \bibinfo{person}{Shaoan Xie}, \bibinfo{person}{Hongning Dai}, \bibinfo{person}{Xiangping Chen}, {and} \bibinfo{person}{Huaimin Wang}.} \bibinfo{year}{2017}\natexlab{}.
\newblock \showarticletitle{An overview of blockchain technology: Architecture, consensus, and future trends}. In \bibinfo{booktitle}{\emph{2017 IEEE international congress on big data (BigData congress)}}. Ieee, \bibinfo{pages}{557--564}.
\newblock


\end{thebibliography}

\end{document}